\documentclass[twocolumn,secnumarabic,amssymb, nobibnotes, aps, pra]{revtex4-1}

\usepackage{framed}
\usepackage{amsfonts}
\usepackage{amsmath}
\usepackage{graphicx}
\usepackage{color}

\usepackage{hyperref}

\begin{document}

\title{Modeling of quasi-phase-matched cavity enhanced second harmonic generation}%
\author{C. Mas Arab\'{i}$^1$, P. Parra-Rivas$^1$, C. Ciret$^2$, S.P. Gorza$^1$, F. Leo$^1$}%
\affiliation{ $^1$OPERA-Photonique CP 194/5, Universit\'e libre de Bruxelles (ULB), Av.F.D. Roosvelt 50, B-1050 Bruxelles, Belgium \\
 $^2$Laboratoire de Photonique d'Angers EA 4464, Universit\'e d'Angers, 2 Bd. Lavoisier, 49000 Angers, France}
\begin{abstract}

We propose a mean-field model to describe second harmonic generation in a resonator made of a material with zincblende crystalline structure. The model is obtained through an averaging of the propagation equations and boundary conditions. It considers the phase-mismatched terms, which act as an effective Kerr effect. We analyze the impact of the different terms on the steady state solutions, highlighting the competition between nonlinearities.

\end{abstract}
\maketitle
\section{Introduction}


Frequency conversion plays an important role in the production of optical sources \cite{fejer_nonlinear_1994}, or in many technological applications in biophotonics \cite{doi:10.1081/ASR-100106156} and quantum information \cite{RevModPhys.84.777}. Nanometric scale waveguides are particularly well suited for nonlinear optics as they allow confining light down to sub-wavelength scales, strongly increasing light-matter interactions.
While many different materials have been used for integrated frequency conversion, the ones with a stronger quadratic response, such as III-V semiconductors and lithium niobate, have recently attracted increasing attention \cite{wang_ultrahigh-efficiency_2018,chang_heterogeneously_2018}. 
Ring resonators in particular have allowed for record conversion efficiencies as both the power and the interactions length are greatly enhanced \cite{Lu:19}. 
Harnessing these resonators for efficient frequency conversion requires to fulfill a phase matching condition \cite{boyd_nonlinear_2008}. 
Quasi-phase-matching (QPM) is attractive because it allows the coupling between fundamental modes, maximizing the effective nonlinearity.
While most often engineered by poling \cite{ilchenko_nonlinear_2004,chen_ultra-efficient_2019} or orientation patterning \cite{Eyres2001}, QPM also naturally occurs in some crystals such as BBO \cite{Lin_APL_2013,Lin:17} and III-V semiconductors \cite{dumeige_whispering-gallery-mode_2006,kuo_4-quasi-phase-matched_2009}.
In the latter, the sign of the effective nonlinearity changes every quarter roundtrip,  because of the $\bar{4}$ symmetry of the material.
The propagation geometry is hence equivalent to that in a poled medium with a $\pi R$ period, where $R$ is the radius of the resonator, and similar nonlinear dynamics is to be expected.
In the case of second harmonic (SH) generation, QPM is known to induce an effective Kerr nonlinearity through cascaded three-wave mixing~\cite{clausen_spatial_1997}.
Because there are mismatched nonlinear interactions, a fraction of the SH wave gets converted back to the pump with a shifted, power dependent, phase~\cite{desalvo_self-focusing_1992}.
This effect allows engineering competing nonlinearities~\cite{bang_engineering_1999}, and can be used for pulse compression \cite{zhou_ultrafast_2012} as well as ultrabroadband light generation \cite{levenius_multistep_2012}. 
However, despite potential applications, the impact of cascaded nonlinearities on the dynamics of QPM resonators is, to the best of our knowledge, still poorly understood.

In this paper, we derive a mean-field model that describes SH generation in a passive quasi-phase-matched III-V-on-insulator cavity. We investigate the impact of cascaded three-wave mixing and find that it can affect the conversion efficiency in some circumstances.

While our analysis focuses on III-V semiconductor rings, we stress that it can readily be generalized to any resonator with a periodic modulation of the nonlinear susceptibility and/or refractive index \cite{Lin_APL_2013,chen_ultra-efficient_2019}.

The paper is organized as follows. In section \ref{sec:2}, we describe light propagation in a bent waveguide and study the impact of the curvature on the nonlinear coefficient. In section \ref{sec:3}, we derive a mean-field equation which models the propagation in a III-V resonator and takes into account the phase-mismatched terms. Section \ref{sec:4} is devoted to the study of stationary solutions. Finally, in section \ref{sec:5}, we summarize our results and discuss their implications on the design of cavity-enhanced SH generation.

\section{Description of the propagation in a resonator}
\label{sec:2}
The resonator under study is sketched in Fig. \ref{fig:figura_1}. It is made of indium gallium phosphide (InGaP), grown in the [010] direction, bonded on silica~\cite{poulvellarie2020second,ciret2020vectorial}. 

The electric field $\mathbf{E}(\mathbf{r}_\bot, \phi,\omega)$ can be expressed as a sum of two modes: 
\begin{align}
\mathbf{E} = a(\phi) \frac{\mathbf{e}_a(\mathbf{r}_\bot,\omega_0)}{\sqrt{N_a}}&e^{i(\beta_a R \phi-\omega_0t)}+ \nonumber \\
+&b(\phi)\frac{\mathbf{e}_b(\mathbf{r}_\bot,2\omega_0)}{\sqrt{N_b}}e^{i(\beta_b R \phi-2\omega_0t)},
\end{align}
where,   $\beta_a$ and $\beta_b$ are the propagation constants of the fundamental wave (FW) and SH, $\omega_0$ is the frequency of the FW, $|a(\phi)|^2$ and $|b(\phi)|^2$ represent the power carried by the FW and SH mode respectively, $\mathbf{e_a}$ and $\mathbf{e_b}$ are the corresponding vector mode distribution of the electric field, and $\phi$ is the azimuthal angle. $N_j$ are the normalization constants provided by the following expression: 

\begin{equation}
\text{Re}\left(\int \mathbf{e}_i(\mathbf{r}_\bot,\omega)\times \mathbf{h}_j^*(\mathbf{r}_\bot,\omega)\cdot \mathbf{dS} \right) = 2N_i\delta_{ij}, 
\end{equation}

where $\delta_{ij}$ is the Kronecker delta and $\mathbf{h}$ is the vector mode distribution of the magnetic field. 
The field amplitudes are governed by the following system of ordinary differential equations:
\begin{align}
R^{-1}\frac{da}{d\phi}&= -\frac{\alpha_a}{2}a + i\kappa^*(\phi)ba^* e^{-i\Delta \beta R \phi} \nonumber \\
R^{-1}\frac{db}{d\phi}&= -\frac{\alpha_b}{2}b + i\kappa(\phi)a^2 e^{i\Delta \beta R \phi} \label{eq:system} 
\end{align}
where  $\Delta\beta=2\beta_a-\beta_b$ corresponds to the wavevector mismatch, $\alpha_a,\alpha_b$ are the loss coefficients associated to the propagation, and $\kappa(\phi)$ is the nonlinear coefficient. To focus on the effects of cascaded three-wave mixing, we neglect higher-order nonlinearities.

The value of $\kappa(\phi)$ is determined by the symmetries of the crystalline structure and the propagation direction. In the case of materials with $\bar{4}3\text{m}$ structure, the only nonzero tensorial element of the electric susceptibility ($\chi_{ijk}^{(2)}$) is $\chi^{(2)}_{xyz}$ with $x\neq y \neq z$. The value of $\chi^{(2)}_{xyz}$ has been measured for indium gallium phosphide to be as high as 220 pm/V around 1550 nm \cite{ueno_second-order_1997}. When the propagation direction is aligned with a  crystallographic axis, the effective nonlinearity ($\kappa$) reads  \cite{ciret2020vectorial}:

\begin{figure}
	\centering
	\includegraphics[scale=0.3]{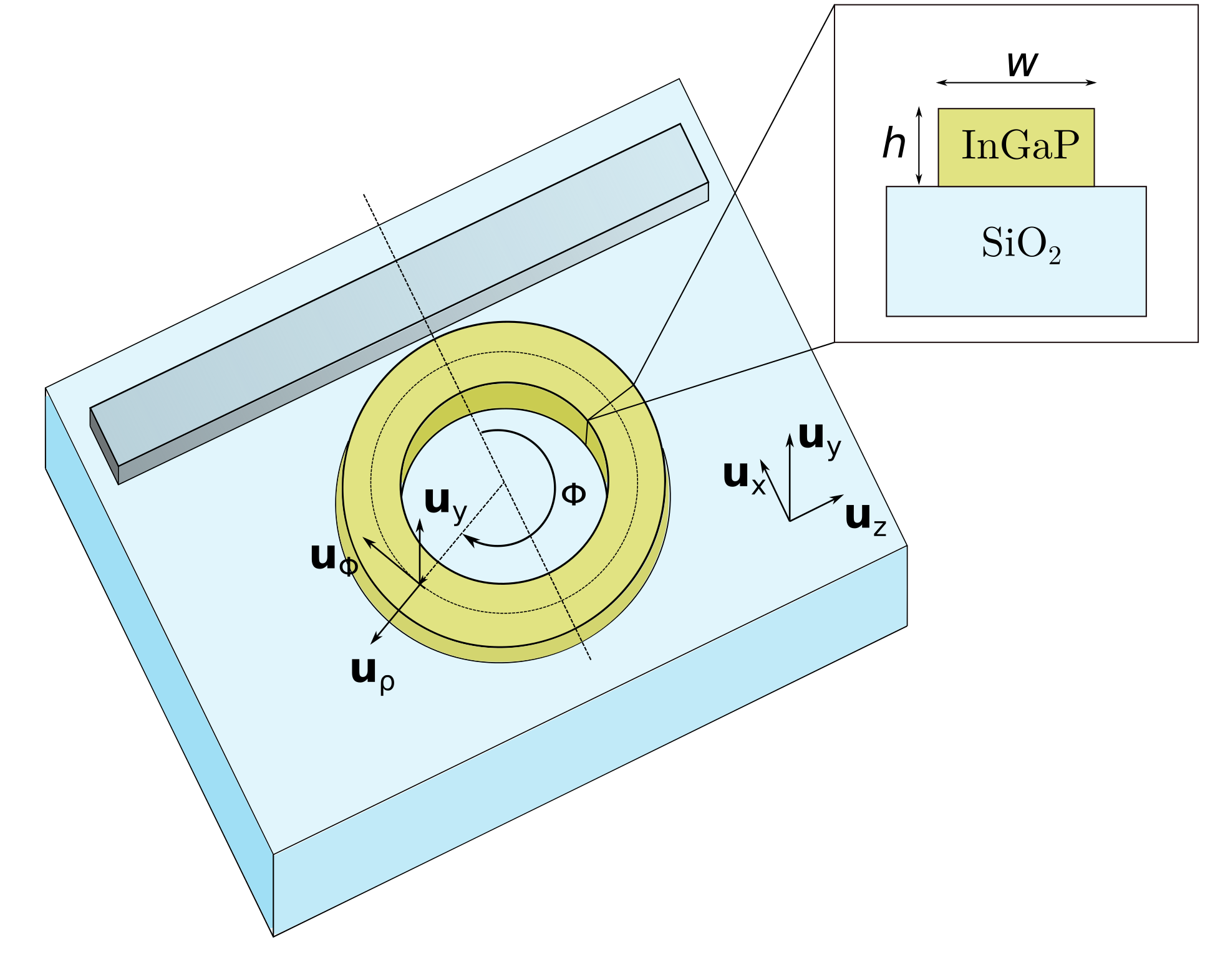}
	\caption{Sketch of the considered resonator. The coordinates $( \mathbf{u}_x,\mathbf{u}_y,\mathbf{u}_z)$ stand for the crystal axis while ($\mathbf{u}_\rho$,$ \mathbf{u}_y$,$\mathbf{u}_\phi$) refer to the cylindrical basis of the resonator. The inset shows the transverse structure of the ring, that is composed of a rectangular waveguide of height $h$ and width $w$.   }
	\label{fig:figura_1}
\end{figure}
\begin{equation}
\kappa = \frac{\epsilon_0\omega_0}{2N_a\sqrt{N_b}}\int_G \chi^{(2)}_{xyz}(e_b^{*x}e_a^{y}e_a^{z}+e_b^{*y}e_a^{x}e_a^{z}+e_b^{*z}e_a^{x}e_a^{y})dS,
\label{eq:5}
\end{equation}
where $e^{k}_{a}(e^{k}_{b})$ is the component of $\mathbf{e}_a(\mathbf{e}_b)$ along the crystal axis $k$,  and the integration is restricted to the indium gallium phosphide cross-section ($G$).
During propagation, the relative orientation between the crystal and the propagation direction changes. Both frames can be related through the transformation $(\mathbf{u}_x,\mathbf{u}_y,\mathbf{u}_z) =( {\mathbf{u}}_\rho\cos(\phi)- {\mathbf{u}}_\phi \sin(\phi), {\mathbf{u}}_y, {\mathbf{u}}_\rho\sin(\phi)+ {\mathbf{u}}_\phi \cos(\phi)) $, where  $( {\mathbf{u}}_x, {\mathbf{u}}_y, {\mathbf{u}}_z)$ are the vectors of the cartesian basis  and  $( {\mathbf{u}}_\rho, {\mathbf{u}}_y, {\mathbf{u}}_\phi)$  the cylindrical ones. For simplicity, we limit ourselves to the most efficient processes, i.e. the ones involving a quasi-TE FW mode and a quasi-TM SH mode~\cite{ciret2020vectorial}. Thus, the first and third terms in Eq. (\ref{eq:5}) can be safely neglected. 

The effective nonlinearity can then be expressed as a sum of two Fourier modes: $\kappa(\phi) = \kappa_+ e^{i2\phi}+\kappa_- e^{-i2\phi}$ \cite{kuo_4-quasi-phase-matched_2009} where:

\begin{equation}
\kappa_\pm \approx \frac{\omega_0\epsilon_0}{4 N_a\sqrt{N_b}}\int_G \chi^{(2)}_{xyz}e_b^{*y}\left(e_a^{\rho}e_a^{\phi} \pm i \frac{(e_a^{\rho})^2+(e_a^{\phi})^2}{2} \right)dS.
\end{equation}

\begin{figure*}
	\centering
	\includegraphics[scale=0.75]{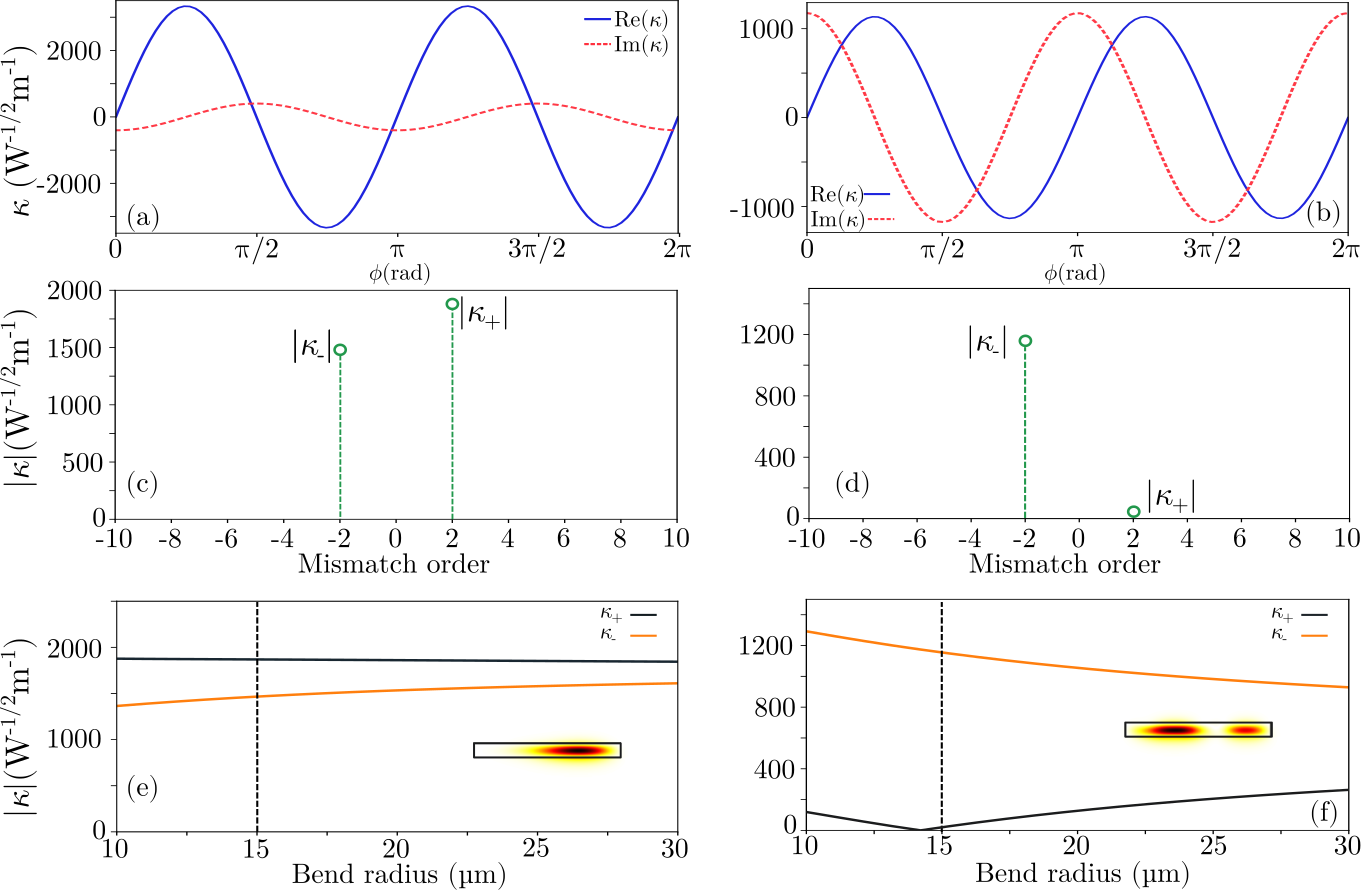}
	\caption{Figs. (a) and (b) show the nonlinear coefficient as a function of $\phi$. Fig. (a) shows $\kappa(\phi)$ when the SH mode is TM$_{00}$. Fig. (b) shows $\kappa(\phi)$ when the SH mode is TM$_{10}$. Figs. (c) and (d) are the coefficients of the Fourier series of the functions represented in Figs. (a) and (b) respectively. Figs. (e) and (f) show  $|\kappa_+|$ and $|\kappa_-|$ as a function of the bending radius for a SH modes TM$_{00}$ (e) and TM$_{10}$ (f). }
	\label{fig:figura_2}
\end{figure*}

The values of $\kappa_{\pm}$ depend on the modal distribution. In Figs. 2(a) and (b) the imaginary and real parts of these nonlinear coefficients are plotted as a function of the angle $\phi$ for two different modes. In  Figs. \ref{fig:figura_2}(c) and (d), we show the coefficients of the Fourier series  of $\kappa(\phi)$. In Fig. \ref{fig:figura_2}(a), (c) the spatial distribution of the SH is the fundamental TM mode, while in Fig. \ref{fig:figura_2}(b) and \ref{fig:figura_2}(d) it is the mode TM$_{10}$. We have calculated $\kappa(\phi)$ by means of the commercial software Lumerical Mode Solutions \cite{MODE}. The waveguide has a width $w=1250~\text{nm}$, a height $h=135~\text{nm}$ and a bend radius of $R=15~\mu \text{m}$. The FW wavelength is fixed to 1550 nm. The values of the nonlinear coefficients are $\kappa_+=-1869i~\text{W}^{-1/2}\text{m}^{-1}$ and $\kappa_-=1466i~\text{W}^{-1/2}\text{m}^{-1}$ when the spatial mode of SH is the TM$_{00}$ and $\kappa_+=20i~\text{W}^{-1/2}\text{m}^{-1}$ and $\kappa_-=1158i~\text{W}^{-1/2}\text{m}^{-1}$ when the SH spatial mode is the TM$_{10}$. It is worth noticing that while in the first case the values of $\kappa_\pm$ are of the same magnitude, when the SH spatial mode is TM$_{10}$, $\kappa_{\pm}$ have very different values \cite{kuo_second-harmonic_2014,dumeige_whispering-gallery-mode_2006}.

Next, we have numerically computed $\kappa_{\pm}$ as a function of the ring radius for the two examples previously considered. Figs. \ref{fig:figura_2}(e) and \ref{fig:figura_2}(f) show $|\kappa_{\pm}|$ as a function of the waveguide bend radius $R$ for the TM$_{00}$ and TM$_{10}$ modes respectively. The insets correspond to the Poynting vector of the SH spatial mode with $R=15~\mu$m. 
In the TM$_{10}$ case, $|\kappa_+|$ can be orders of magnitude smaller than $|\kappa_-|$. For radii close to $15~\mu$m, we expect phase mismatched SHG to impact the dynamics. To investigate this regime of cascaded non-linearities, we derive below a mean-field model that includes the phase-mismatched terms.


\section{Derivation of the model}
\label{sec:3}

Our starting point is the system of equations Eqs. (\ref{eq:system}) and the following boundary condition equations: 
\begin{align}
&a^{(m+1)}(0)=\sqrt{1-\theta^2} a^{(m)}(2\pi)e^{i\beta_aL }+\theta a_{in} \nonumber \\
&b^{(m+1)}(0)=\sqrt{1-\theta^2} b^{(m)}(2\pi)e^{i\beta_bL},
\label{eq:BC}
\end{align}
where $a^{(m)}, b^{(m)}$ are the field amplitudes at the $m^{th}$ roundtrip, $|a_{in}|^2$ is the input power, $L=2\pi R$ is the resonator length, and  $\theta$ is the field transmission coefficient of the coupler. These conditions link the output of the $m^{th}$ roundtrip with the input of the $(m+1)^{th}$ one. To further proceed, boundary conditions are injected into the evolution equations. A similar approach has been already employed to generalize the Lugiato-Lefever model of passive fiber cavities \cite{conforti_multi-resonant_2017} and microresonators \cite{xue_second-harmonic-assisted_2017}. This method consists in unfolding the cavity and modeling it as a waveguide with periodic localized gain and losses. Mathematically, these conditions can be expressed by making use of a Dirac delta comb, that we include in the right-hand side of Eqs. \eqref{eq:system}. We expand both propagation constants around the closest resonances:  $\beta_{(a,b)}=(2\pi n_{(a,b)} -\delta_{(a,b)})/L$ where $n_{(a,b)}$ are integer numbers. In addition, we perform a phase-rotation $a\rightarrow ae^{i\frac{\delta_a}{2\pi}\phi}$, and $b\rightarrow b e^{i\frac{\delta_b}{2\pi}\phi}$. Therefore, the evolution equations read:
\begin{align}
&R^{-1}\frac{da}{d\phi}+\left(i\frac{\delta_a}{L}+\frac{\alpha_a}{2}\right)a-i\kappa^*(\phi)ba^*e^{-i(2n_a-n_b)\phi}= \nonumber\\
& \qquad \qquad \qquad \qquad    \left(-\frac{\alpha_cLa}{2}+\theta a_{in}\right)\sum_m \delta(R\phi-  m L)\nonumber \\
& R^{-1}\frac{db}{d\phi}+\left(i\frac{\delta_b}{L}+\frac{\alpha_b}{2}\right)b-i\kappa(\phi)a^2e^{i(2n_a-n_b)\phi}=  \nonumber\\
& \qquad \qquad \qquad  \qquad \qquad  -\frac{\alpha_c L b}{2}\sum_m \delta(R\phi-mL ), 
\label{eq:A2}
\end{align}
where $\alpha_c=\theta^2L^{-1}$. These equations are similar to Eqs. \eqref{eq:system}, but forced by a Dirac delta comb. The forcing that appears in Eqs. (\ref{eq:A2}) models the periodic gain and losses described by the boundary conditions of the resonator. The Dirac delta comb can be expressed as a Fourier series by making use of the Poisson resummation identity: 
\begin{equation}
\sum_m \delta(R\phi- m L)=\frac{1}{L}\sum_m e^{i m \phi}.
\label{eq:Poisson}
\end{equation}  
Proceeding in this way, and employing Eq. \eqref{eq:Poisson},  Eqs. \eqref{eq:A2} become: 
\begin{align}
&R^{-1}\frac{da}{d\phi}+\left(i\frac{\delta_a}{L}+\frac{\alpha_a}{2}\right)a-i\kappa^*(\phi)ba^*e^{-i(2n_a-n_b)\phi}= \nonumber\\
& \qquad \qquad \qquad \qquad \qquad \qquad   \left(-\frac{\alpha_c a}{2}+\frac{\theta}{L}a_{in}\right)\sum_m e^{i m \phi}\nonumber \\
& R^{-1}\frac{db}{d\phi}+\left(i\frac{\delta_b}{L}+\frac{\alpha_b}{2}\right)b-i\kappa(\phi)a^2e^{i(2n_a-n_b)\phi}= \nonumber \\
& \qquad \qquad \qquad \qquad \qquad \qquad  \quad \quad \quad-\frac{\alpha_c b}{2}\sum_m e^{i m \phi}.
\label{eq:Eq9}
\end{align}
Without loss of generality, the process involving $\kappa_+$ is considered to be quasi-phase-matched, i.e. $2n_a-n_b=-2$.  However, the results that we will obtain can be easily generalized to  $2n_a-n_b=2$, or for any poled resonator where QPM is verified.  Thus, the evolution equations become:
\begin{align}
&R^{-1}\frac{da}{d\phi}+\left(i\frac{\delta_a}{L}+\frac{\alpha_p}{2}\right)a-i(\kappa_+^*+\kappa_-^*e^{+ i4\phi })ba^*= \nonumber\\
& \qquad \qquad \qquad \qquad \qquad \qquad   \left(-\frac{\alpha_c a}{2}+\frac{\theta}{L}a_{in}\right)\sum_m e^{im \phi}\nonumber \\
& R^{-1}\frac{db}{d\phi}+\left(i\frac{\delta_b}{L}+\frac{\alpha_p}{2}\right)b-i(\kappa_++\kappa_-e^{- i4\phi })a^2= \nonumber \\
& \qquad ~~ \qquad \qquad \qquad \qquad \qquad \qquad -\frac{\alpha_c b}{2}\sum_m e^{i m \phi}. 
\label{eq:pre_averaged}
\end{align}
The detunings $\delta_a$ and $\delta_b$ are linked through the relation $\delta_b=2\delta_a=2\delta_0$  if $\Delta\beta=-2R^{-1}$. 
As a first order approximation, we can neglect  complex exponentials in the left-hand side of   Eqs. (\ref{eq:pre_averaged}). However, this simplification, also known as the fast-rotating wave (FRW) approximation, may not be valid for large values of $|\kappa_-|$. In order to derive a model that also describes the regimes where $\kappa_-$ has a non-negligible contribution, we employ the averaging method described in Ref. \cite{kivshar_kinks_1994}. We express $a(\phi)$ and $b(\phi)$ as Fourier series $(a,b)=\sum_k (a_k(\phi),b_k(\phi))e^{ik\phi}$, where each coefficient $k$ relates to the longitudinal mode $N$ of the cavity as $N_{a,b}=n_{a,b}+k$,  and substitute them in Eqs. (\ref{eq:pre_averaged}). Then, by collecting the terms that oscillate at the same frequency, considering critical coupling, and equal loss for both modes $\alpha_c=\alpha_a=\alpha_b=\alpha$, we find: 
\begin{widetext}
\begin{align}
&\frac{dA_k}{d\bar{\phi}}+\left(i(\Delta+2\mathcal{F} k)+\frac{1}{2}\right)A_k-i\sum_m\left(\frac{\kappa_-^*}{\kappa_+^*}A_{m+4-k}^*B_m+A_{m-k}^*B_m\right)=-\frac{1}{2}\sum_m A_m+\rho, \nonumber \\
&\frac{dB_k}{d\bar{\phi}}+\left(2i(\Delta+\mathcal{F} k)+\frac{1}{2}\right) B_k-i\sum_m\left(\frac{\kappa_-}{\kappa_+}A_{k+4-m}A_m+A_{m-k} A_m\right)=-\frac{1}{2}\sum_m B_m  ,
\label{eq:A3}
\end{align}
\end{widetext}
where we introduced the finesse $\mathcal{F}=\pi(\alpha L)^{-1}$ and the following normalization: $\Delta=\delta\alpha L^{-1} $,  $\rho=\kappa_+La_{in}(\mathcal{F}/\pi)^{3/2}$, $(A_k,B_k)=\alpha\kappa_+^{-1}(a_k,b_k)$, $\bar{\phi}=R  \alpha^{-1} \phi$.
We now have a system of equations for each longitudinal mode of the cavity. These modes interact via the nonlinear terms and the periodic losses induced by the coupling with the bus waveguide. In this expansion, the order $k=0$ represents the averaged dynamics over one roundtrip. Note that all the coefficients of the system do not have the same order of magnitude, while $\Delta$ and $\rho$ are close to 1, $\mathcal{F}$ may reach values of hundreds for high-finesse resonators.
This difference allows to make use of a multi-scale expansion when $k\neq 0$. By writing $(A_k,B_k)=\sum_l (A_k^{(l)},B_k^{(l)})\epsilon^l$, where the multi-scale parameter is $\epsilon=\mathcal{F}^{-1}$, an infinite hierarchy of algebraic equations  is obtained: 
\begin{align}
\epsilon^{-1}:\quad A^{(0)}_{-4} = 0~,~B^{(0)}_{4} = 0 \\
\epsilon^{0}:\quad A_{4}^{(1)}=\frac{1}{8}\left(\frac{\kappa_-^*}{\kappa_+^*}A_0^*B_0+\left(\rho-\frac{A_0}{2}\right)\right), \nonumber \\
\quad B_{-4}^{(1)}=-\frac{1}{8}\left(\frac{\kappa_-}{\kappa_+}A_0^2+\frac{B_0}{2}\right),
\end{align}
where we only show the cases where there is a coupling with the order $k= 0$ at $\epsilon^0$. For low values of $\rho$ and high values of $|\kappa_-/\kappa_+|$, we can make the approximation:
\begin{equation}
A_{4}^{(1)}=\frac{\kappa_-^*}{8\kappa_+^*}A_0^*B_0\quad ,
\quad B_{-4}^{(1)}=-\frac{\kappa_-}{8\kappa_+}A_0^2.
\label{eq:A4Bm4}
\end{equation}
These equations relate the longitudinal mode $k=0$ with the modes $k=\pm4$. Therefore, the evolution of the dimensional fields $a(\phi)$ and $b(\phi)$ can be expressed as: 
\begin{align}
&a(\phi)=a_0+\frac{\kappa^*_-R}{4} a^*_0b_0e^{i4 \phi} \nonumber\\
&b(\phi)=b_0-\frac{\kappa_-R}{4} a_0^2e^{-i4\phi},
\label{eq:Field_complet}
\end{align}
where $a_0$ and $b_0$ are governed by the following differential equations:
\begin{align}
&R^{-1}\frac{da_0}{d\phi}+\left(i\frac{\delta_0}{L}+\alpha\right)a_0-i\kappa_+^*b_0a_0^*- \nonumber\\
& \qquad \qquad \qquad \qquad  \qquad \qquad  i\gamma (|a_0|^2-|b_0|^2)a_0= \frac{\theta}{L}a_{in} \nonumber \\
&R^{-1}\frac{db_0}{d\phi}+\left(i\frac{2\delta_0}{L}+\alpha\right)b_0-i\kappa_+ a^2_0+2i\gamma|a_0|^2 b_0=0,
\label{eq:averaged_dimensional}
\end{align}
with $\gamma =- |\kappa_- |^2 L /(8\pi )$. Interestingly, higher-order wave-mixing terms, akin to a third-order nonlinearity, appear in Eq.~\eqref{eq:averaged_dimensional}. Note that this nonlinearity does not exactly correspond to a pure Kerr interaction, since there is no self-phase modulation term in the second equation. In addition, the self- and cross-phase modulation terms in the first equation are of opposite signs. The magnitude of the third order nonlinearity corresponds to the well-known cascading limit, which is $\gamma= |\kappa_-|^2/(2\Delta \beta)$ \cite{clausen_spatial_1997}. 
\begin{figure}
	\includegraphics[scale=0.9]{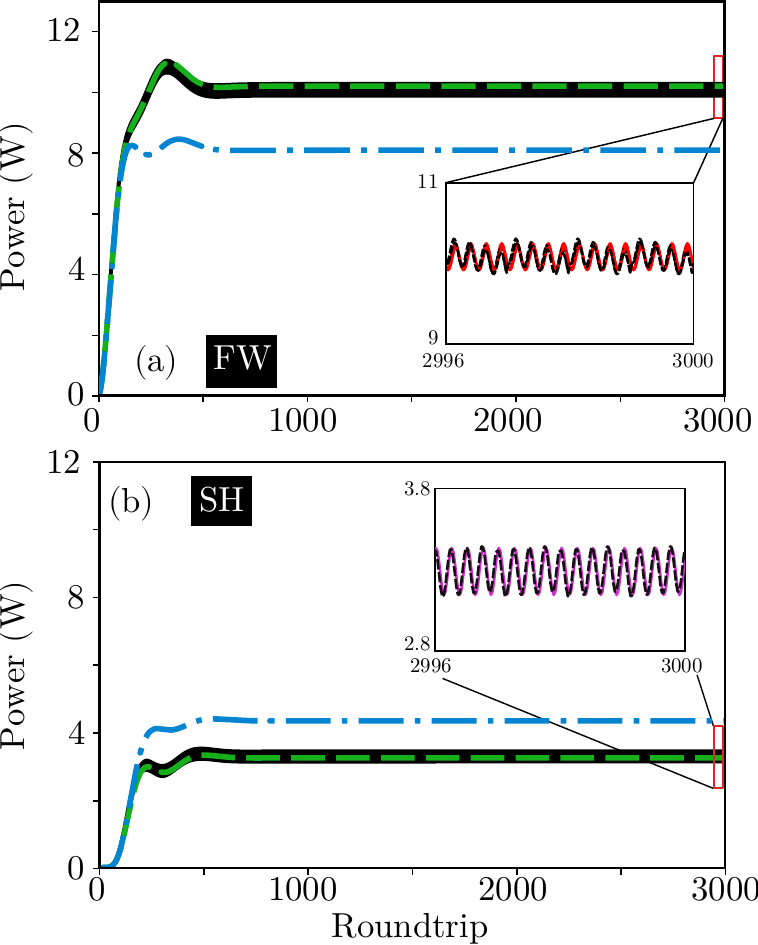}
	\caption{Evolution of power in the FW (a) and SH (b) modes. Black (solid) line corresponds to the map, blue (point-dashed) line corresponds to Eqs. (\ref{eq:averaged_dimensional}) with $\gamma=0$, and green (dashed) line is the solution of Eqs. (\ref{eq:averaged_dimensional}) with $\gamma\neq$0. The parameters are: $\Delta=$-0.4, $|a_{in}|^2=0.2~\text{W}$, $|\kappa_+|=38$ W$^{-1/2}$m$^{-1}$, $|\kappa_-|=1101$ W$^{-1/2}$m$^{-1}$, $R$=14.5 $\mu$m.  }
	\label{fig:figura_3}
\end{figure}
\begin{figure*}
	\includegraphics[scale=0.8]{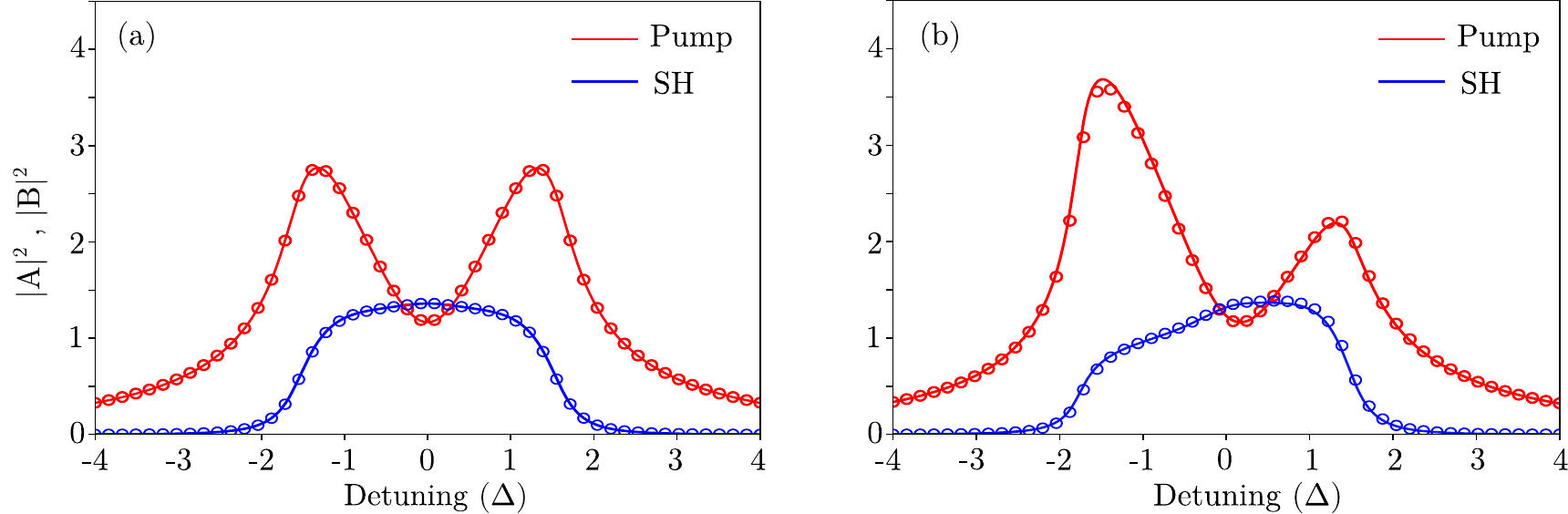}
	\caption{ Comparative between the steady states obtained with the exact map (circles) and the averaged equations  Eqs.\eqref{eq:adimensional} (blue line for SH and red line for the FW).  (a) The phase-matched term mode is TM$_{00}$. Coefficients are $\kappa_-=$ 1305i W$^{-1/2}$m$^{-1}$, $\kappa_+$=-1613i W$^{-1/2}$m$^{-1}$, $\bar{\gamma}=-3\times 10 ^{-4}$ and $h=122$ nm.  (b) The phase-matched mode is TM$_{10}$.  Coefficients are $\kappa_-=$ 1083i W$^{-1/2}$ m$^{-1}$, $\kappa_+$=57i W$^{-1/2}$m$^{-1}$, $\bar{\gamma}$ = -0.16 and $h=130$ nm. }
	\label{fig:figura_4}
\end{figure*}

In order to see the effect of phase-mismatched terms and compare results to the outcome of a model where the FRW approximation is used, we performed some numerical simulations. We considered a resonator with $\mathcal{F}=300$, pumped in the TE$_{00}$ mode with $|a_{in}|^2=200~\text{mW}$. We supposed a critically coupled resonator. The dimensions of the resonator were chosen such that the process involving $\kappa_+$ verified QPM at $\lambda_p=1550~\text{nm}$ with the SH spatial mode TM$_{10}$: $h=130~\text{nm}$, $w=1250~\text{nm}$ and $R=14.5~\mu\text{m}$. With these values, we find $\kappa_+=38i~\text{W}^{-1/2}\text{m}^{-1}$ and $\kappa_-=1101i~\text{W}^{-1/2}\text{m}^{-1}$. Furthermore, here we chose a normalized detuning $\Delta=-0.4$. In Figs.  \ref{fig:figura_3}(a) and (b) we show the intracavity power evolution of the FW and SH fields. The dark solid line corresponds to the map, Eqs. (\ref{eq:system}) and (\ref{eq:BC}). The fast oscillations are due to the phase-mismatched terms and are clearly appreciable in the insets. The blue dashed-dotted line corresponds to Eqs. (\ref{eq:averaged_dimensional}) after applying the FRW approximation, i.e. setting $\gamma=0$. From the figure it is clear that the FRW approximation is not valid in the considered regime because the mean value of the stationary states is significantly different. The green dashed line is the numerical solution of Eqs. (\ref{eq:averaged_dimensional}). In the insets of Fig. \ref{fig:figura_3}, we display a zoom of the last roundtrips.  The result obtained through Eqs. (\ref{eq:Field_complet}) are plotted in dashed red in Fig. \ref{fig:figura_3}(a) and in dashed magenta in Fig. \ref{fig:figura_3}(b). The numerical results are shown in black for comparison.
The value of $a_0$ and $b_0$ employed in the analytical expression is obtained from the averaged value of the fields over one roundtrip. We can see that both curves are nearly superimposed. 
We find excellent agreement between our model and the map equations  (\ref{eq:system}) and (\ref{eq:BC}), validating the use of a mean-field equation to describe the competition between the phase-matched terms, driving the frequency conversion, and an effective Kerr nonlinearity imposed by the phase-mismatched terms.

\section{Stationary solutions}
\label{sec:4}

Next, we analyze the impact of the phase-mismatched terms on the stationary solutions of the system. In normalized form,  Eqs. (\ref{eq:averaged_dimensional}) read:  
\begin{align}
&\frac{dA_0}{d\bar{\phi}}+(1+i\Delta)A_0-i(\bar{\gamma}(|A_0|^2-|B_0|^2)A_0+B_0A^*)=\rho \nonumber \\
&\frac{dB_0}{d\bar{\phi}}+(1+i2\Delta)B_0-i(A_0^2-2\bar{\gamma}|A_0|^2B_0)=0
\label{eq:adimensional}
\end{align}
where $\bar{\gamma} =\text{sign}(\Delta \beta)|\kappa_-|^2/(8|\kappa_+|^2\mathcal{F})$. The normalized coefficient $\bar{\gamma}$ shows the relevance of phase-mismatched terms. Note that $\bar{\gamma}$ has a non-negligible value when $|\kappa_\pm|$ have different order of magnitude. More precisely, the coefficient $\bar{\gamma}$ confirms that mismatched terms are relevant when $|\kappa_-|\gg |\kappa_+|$. In order to study the impact of $\bar{\gamma}$ and verify the validity of the mean-field equations for different values of the detuning, we compared the steady state solutions obtained with the map and with Eqs. (\ref{eq:adimensional}). In order to get rid of eventual oscillations of the map solutions, we calculated the mean value of the intensity over one roundtrip. The stationary states of Eqs. (\ref{eq:adimensional}) were found by imposing $\dot{A}=\dot{B}=0$ and making use of numerical continuation by means of the free distribution program AUTO \cite{Doedel07auto-07p:continuation}. The value of $|a_{in}|$ is chosen such that $\rho$ is equal to 2.34 in both cases. This value of $\rho$ is below the threshold for bistablity and self-pulsing \cite{drummond_non-equilibrium_1980}. The resonator dimensions are chosen so that SH generation is quasi-phase-matched at $\lambda_p=1550~\text{nm}$ with $R=15.25~\mu\text{m}$ and $w=1250~\text{nm}$. The finesse is set to $\mathcal{F}=300$.  

The first case corresponds to a resonator where QPM is verified for the TM$_{00}$ mode. The value of the height is $h=122 ~\text{nm}$. The values of the nonlinear coefficients are $\kappa_+=1613i~\text{W}^{-1/2}\text{m}^{-1}$ and  $\kappa_-=1305i~\text{W}^{-1/2}\text{m}^{-1}$.  The value of  $\bar{\gamma}=-3\times10^{-4}$ and thus, the impact of mismatched terms is negligible. The intracavity power of the steady states as a function of $\Delta$ is displayed in Fig. \ref{fig:figura_4}(a). Within this limit, resonances are symmetric with respect to the detuning and the maximum of intracavity power for second harmonic is found for zero detuning \cite{villois_soliton_2019,hansson_quadratic_2018}. 

Next, we studied the case where the SH propagates in a TM$_{10}$ mode. The height is $h=130 ~\text{nm}$. The values of the nonlinear coefficients are $\kappa_+=57i~\text{W}^{-1/2}\text{m}^{-1}$, and  $\kappa_-=1083i~\text{W}^{-1/2}\text{m}^{-1}$. In this case, $\bar{\gamma} = -0.16$, the impact of cascaded nonlinearities can be clearly seen on the resonances. They become asymmetric, akin to those found when the Kerr nonlinearity is included~\cite{villois_soliton_2019}.

We stress that the two configurations shown in Fig. \ref{fig:figura_4} are very different. The conversion is much weaker in the latter case because the resonator is engineered such that the weaker nonlinear mode ($\kappa_+$) is quasi-phase-matched so as to maximize the impact of cascaded nonlinearities. This difference can be appreciated by calculating the conversion efficiency ($\eta={\theta |b_{max}|^2}/{|a_{in}|^4}$) for the two configurations.
It is as high as {$\eta=970,000~\text{\%/W}$} in the first case, which is of the same order of magnitude as the state of the art~\cite{Lu:19}. However, in the other case, we find a significantly lower conversion efficiency of $\eta=1,150~\text{\%/W}$.

Our results hence suggest that the impact of cascaded nonlinearities is negligible as long as the larger nonlinear mode is quasi-phase-matched, as indicated from the normalized parameter $\bar{\gamma} = \text{sign}(\Delta \beta)|\kappa_-|^2/(8|\kappa_+|^2\mathcal{F})$. 
Yet, we showed that modal phase matching allows to engineer competing nonlinearities, and 
expect the mismatched terms to have a significant impact on the known oscillatory and modulation instabilities arising in cavity-enhanced second harmonic generation \cite{drummond_non-equilibrium_1980,leo_walk-off-induced_2016}.

\section{Conclusions}
\label{sec:5}

We have analyzed quasi-phase-matched SH generation in a ring resonator made of a semiconductor with zincblende structure, and hence a $\bar{4}3\text{m}$ symmetry. Starting from the boundary conditions and propagation equations,
we derived a mean-field model that takes into account the phase-mismatched processes.  We showed that they act as an effective third-order nonlinearity which plays a fundamental role for certain configurations. Our analysis can readily be generalized to resonators made of poled materials.

\section*{Acknowledgements}
This work was supported by funding from the European Research Council (ERC) under the European Union's Horizon 2020 research and innovation programme (grant agreement Nos 757800). P. P-R acknowledges the support from the Fonds de la Recherche Scientifique F.R.S.-FNRS.

\bibliographystyle{ieeetr}
\bibliography{biblioteca}

\begin{thebibliography}{10}

\bibitem{fejer_nonlinear_1994}
M.~M. Fejer, ``Nonlinear {Optical} {Frequency} {Conversion},'' {\em Physics
  Today}, vol.~47, pp.~25--32, May 1994.

\bibitem{doi:10.1081/ASR-100106156}
W.~Petrich, ``Mid-infrared and raman spectroscopy for medical diagnostics,''
  {\em Applied Spectroscopy Reviews}, vol.~36, no.~2-3, pp.~181--237, 2001.

\bibitem{RevModPhys.84.777}
J.-W. Pan, Z.-B. Chen, C.-Y. Lu, H.~Weinfurter, A.~Zeilinger, and
  M.~\ifmmode~\dot{Z}\else \.{Z}\fi{}ukowski, ``Multiphoton entanglement and
  interferometry,'' {\em Rev. Mod. Phys.}, vol.~84, pp.~777--838, May 2012.

\bibitem{wang_ultrahigh-efficiency_2018}
C.~Wang, C.~Langrock, A.~Marandi, M.~Jankowski, M.~Zhang, B.~Desiatov, M.~M.
  Fejer, and M.~Lončar, ``Ultrahigh-efficiency wavelength conversion in
  nanophotonic periodically poled lithium niobate waveguides,'' {\em Optica},
  vol.~5, p.~1438, Nov. 2018.

\bibitem{chang_heterogeneously_2018}
L.~Chang, A.~Boes, X.~Guo, D.~T. Spencer, M.~J. Kennedy, J.~D. Peters,
  N.~Volet, J.~Chiles, A.~Kowligy, N.~Nader, D.~D. Hickstein, E.~J. Stanton,
  S.~A. Diddams, S.~B. Papp, and J.~E. Bowers, ``Heterogeneously {Integrated}
  {GaAs} {Waveguides} on {Insulator} for {Efficient} {Frequency}
  {Conversion},'' {\em Laser \& Photonics Reviews}, vol.~12, p.~1800149, Oct.
  2018.

\bibitem{Lu:19}
J.~Lu, J.~B. Surya, X.~Liu, A.~W. Bruch, Z.~Gong, Y.~Xu, and H.~X. Tang,
  ``Periodically poled thin-film lithium niobate microring resonators with a
  second-harmonic generation efficiency of 250,000\%/w,'' {\em Optica}, vol.~6,
  pp.~1455--1460, Dec 2019.

\bibitem{boyd_nonlinear_2008}
R.~Boyd, {\em Nonlinear {Optics}}.
\newblock Elsevier, third edition~ed., 2008.

\bibitem{ilchenko_nonlinear_2004}
V.~S. Ilchenko, A.~A. Savchenkov, A.~B. Matsko, and L.~Maleki, ``Nonlinear
  {Optics} and {Crystalline} {Whispering} {Gallery} {Mode} {Cavities},'' {\em
  Physical Review Letters}, vol.~92, p.~043903, Jan. 2004.

\bibitem{chen_ultra-efficient_2019}
J.-Y. Chen, Z.-H. Ma, Y.~M. Sua, Z.~Li, C.~Tang, and Y.-P. Huang,
  ``Ultra-efficient frequency conversion in quasi-phase-matched lithium niobate
  microrings,'' {\em Optica}, vol.~6, p.~1244, Sept. 2019.

\bibitem{Eyres2001}
L.~A. Eyres, P.~J. Tourreau, T.~J. Pinguet, C.~B. Ebert, J.~S. Harris, M.~M.
  Fejer, L.~Becouarn, B.~Gerard, and E.~Lallier, ``All-epitaxial fabrication of
  thick, orientation-patterned gaas films for nonlinear optical frequency
  conversion,'' {\em Applied Physics Letters}, vol.~79, no.~7, pp.~904--906,
  2001.

\bibitem{Lin_APL_2013}
G.~Lin, J.~U. F\"urst, D.~V. Strekalov, and N.~Yu, ``Wide-range cyclic phase
  matching and second harmonic generation in whispering gallery resonators,''
  {\em Applied Physics Letters}, vol.~103, no.~18, p.~181107, 2013.

\bibitem{Lin:17}
G.~Lin, A.~Coillet, and Y.~K. Chembo, ``Nonlinear photonics with high-{Q}
  whispering-gallery-mode resonators,'' {\em Adv. Opt. Photon.}, vol.~9,
  pp.~828--890, Dec 2017.

\bibitem{dumeige_whispering-gallery-mode_2006}
Y.~Dumeige and P.~F\'eron, ``Whispering-gallery-mode analysis of phase-matched
  doubly resonant second-harmonic generation,'' {\em Physical Review A},
  vol.~74, p.~063804, Dec. 2006.

\bibitem{kuo_4-quasi-phase-matched_2009}
P.~S. Kuo, W.~Fang, and G.~S. Solomon, ``$\bar{4}$-quasi-phase-matched
  interactions in {GaAs} microdisk cavities,'' {\em Optics Letters}, vol.~34,
  p.~3580, Nov. 2009.

\bibitem{clausen_spatial_1997}
C.~B. Clausen, O.~Bang, and Y.~S. Kivshar, ``Spatial {Solitons} and {Induced}
  {Kerr} {Effects} in {Quasi}-{Phase}-{Matched} {Quadratic} {Media},'' {\em
  Physical Review Letters}, vol.~78, pp.~4749--4752, June 1997.

\bibitem{desalvo_self-focusing_1992}
R.~DeSalvo, H.~Vanherzeele, D.~J. Hagan, M.~Sheik-Bahae, G.~Stegeman, and E.~W.
  Van~Stryland, ``Self-focusing and self-defocusing by cascaded second-order
  effects in {KTP},'' {\em Optics Letters}, vol.~17, p.~28, Jan. 1992.

\bibitem{bang_engineering_1999}
O.~Bang, C.~B. Clausen, P.~L. Christiansen, and L.~Torner, ``Engineering
  competing nonlinearities,'' {\em Optics Letters}, vol.~24, p.~1413, Oct.
  1999.

\bibitem{zhou_ultrafast_2012}
B.~B. Zhou, A.~Chong, F.~W. Wise, and M.~Bache, ``Ultrafast and
  {Octave}-{Spanning} {Optical} {Nonlinearities} from {Strongly}
  {Phase}-{Mismatched} {Quadratic} {Interactions},'' {\em Physical Review
  Letters}, vol.~109, p.~043902, July 2012.

\bibitem{levenius_multistep_2012}
M.~Levenius, M.~Conforti, F.~Baronio, V.~Pasiskevicius, F.~Laurell,
  C.~De~Angelis, and K.~Gallo, ``Multistep quadratic cascading in broadband
  optical parametric generation,'' {\em Optics Letters}, vol.~37, p.~1727, May
  2012.

\bibitem{poulvellarie2020second}
N.~Poulvellarie, U.~Dave, K.~Alexander, C.~Ciret, M.~Billet, C.~{Mas Arabi},
  F.~Raineri, S.~Combrie, A.~D. Rossi, G.~Roelkens, S.-P. Gorza, B.~Kuyken, and
  F.~Leo, ``Second harmonic generation enabled by longitudinal electric field
  components in photonic wire waveguides,'' {\em arXiv:2001.01709}, 2020.

\bibitem{ciret2020vectorial}
C.~Ciret, K.~Alexander, N.~Poulvellarie, M.~Billet, C.~{Mas Arabi}, B.~Kuyken,
  S.-P. Gorza, and F.~Leo, ``Full vectorial modeling of second harmonic
  generation in iii-v-on-insulator nanowires,'' {\em arXiv:2001.02210}, 2020.

\bibitem{ueno_second-order_1997}
Y.~Ueno, V.~Ricci, and G.~I. Stegeman, ``Second-order susceptibility of
  {Ga$_{0.5}$In$_{0.5}$P} crystals at 1.5 $\mu$m and their feasibility for
  waveguide quasi-phase matching,'' {\em J. Opt. Soc. Am. B}, vol.~14,
  pp.~1428--1436, Jun 1997.

\bibitem{MODE}
``https://www.lumerical.com/products/mode/.''
  \url{http://www.lumerical.com/products/mode/}.

\bibitem{kuo_second-harmonic_2014}
P.~S. Kuo, J.~Bravo-Abad, and G.~S. Solomon, ``Second-harmonic generation using
  $\bar{4}$-quasi-phasematching in a {GaAs} whispering-gallery-mode
  microcavity,'' {\em Nature Communications}, vol.~5, p.~3109, Dec. 2014.

\bibitem{conforti_multi-resonant_2017}
M.~Conforti and F.~Biancalana, ``Multi-resonant {Lugiato}–{Lefever} model,''
  {\em Optics Letters}, vol.~42, p.~3666, Sept. 2017.

\bibitem{xue_second-harmonic-assisted_2017}
X.~Xue, F.~Leo, Y.~Xuan, J.~A. Jaramillo-Villegas, P.-H. Wang, D.~E. Leaird,
  M.~Erkintalo, M.~Qi, and A.~M. Weiner, ``Second-harmonic-assisted four-wave
  mixing in chip-based microresonator frequency comb generation,'' {\em Light:
  Science \& Applications}, vol.~6, pp.~e16253--e16253, Apr. 2017.

\bibitem{kivshar_kinks_1994}
Y.~S. Kivshar, N.~Gr{\o}nbech-Jensen, and R.~D. Parmentier, ``Kinks in the
  presence of rapidly varying perturbations,'' {\em Physical Review E},
  vol.~49, pp.~4542--4551, May 1994.

\bibitem{Doedel07auto-07p:continuation}
E.~J. Doedel, T.~F. Fairgrieve, B.~Sandstede, A.~R. Champneys, Y.~A. Kuznetsov,
  and X.~Wang, ``Auto-07p: Continuation and bifurcation software for ordinary
  differential equations,'' 2007.

\bibitem{drummond_non-equilibrium_1980}
P.~Drummond, K.~McNeil, and D.~Walls, ``Non-equilibrium {Transitions} in
  {Sub}/{Second} {Harmonic} {Generation},'' {\em Optica Acta: International
  Journal of Optics}, vol.~27, pp.~321--335, Mar. 1980.

\bibitem{villois_soliton_2019}
A.~Villois and D.~V. Skryabin, ``Soliton and quasi-soliton frequency combs due
  to second harmonic generation in microresonators,'' {\em Optics Express},
  vol.~27, p.~7098, Mar. 2019.

\bibitem{hansson_quadratic_2018}
T.~Hansson, P.~Parra-Rivas, M.~Bernard, F.~Leo, L.~Gelens, and S.~Wabnitz,
  ``Quadratic {Soliton} {Combs} in {Doubly}-{Resonant} {Second}-{Harmonic}
  {Generation},'' {\em Optics Letters}, vol.~43, p.~6033, Dec. 2018.

\bibitem{leo_walk-off-induced_2016}
F.~Leo, T.~Hansson, I.~Ricciardi, M.~De~Rosa, S.~Coen, S.~Wabnitz, and
  M.~Erkintalo, ``Walk-{Off}-{Induced} {Modulation} {Instability}, {Temporal}
  {Pattern} {Formation}, and {Frequency} {Comb} {Generation} in
  {Cavity}-{Enhanced} {Second}-{Harmonic} {Generation},'' {\em Physical Review
  Letters}, vol.~116, Jan. 2016.

\end{thebibliography}

\end{document}